\documentclass[showpacs,preprintnumbers,onecolumn,superscriptaddress,aps,nofootinbib]{revtex4}
\usepackage{amssymb}
\usepackage[centertags]{amsmath}
\usepackage{txfonts}
\usepackage{epsfig}
\usepackage{bm}
\usepackage{color}
\usepackage{graphicx,graphics}
\usepackage{multirow}
\usepackage{float}
\usepackage{ulem}
\usepackage[pdfstartview=FitH]{hyperref}
\hypersetup{colorlinks=true, citecolor=blue,linkcolor=blue,filecolor=black,urlcolor=blue}
\allowdisplaybreaks[2]

\begin{document}

\title{Yield correlations and $p_T$ dependence of charm hadrons in Pb + Pb collisions at $\sqrt{s_{NN}}= 2.76$ TeV}

\author{Rui-qin Wang}
\affiliation{School of Physics, Shandong University, Jinan, Shandong 250100, China}

\author{Jun Song}
\affiliation{Department of Physics, Jining University, Shandong 273155, China}

\author{Feng-lan Shao}
\email {shaofl@mail.sdu.edu.cn}
\affiliation{Department of Physics, Qufu Normal University, Shandong 273165, China}

\begin{abstract}

Recently the ALICE Collaboration has published a few data of charm hadrons in Pb + Pb collisions at the CERN Large Hadron Collider (LHC).
We extend the quark combination to the charm sector and point out that the measurement of charm hadron yields can provide important
insights into charm quark hadronization mechanism and properties of the hot and dense matter produced in high energy reactions.
We propose several types of yield ratios, e.g. $D^{*+}/D^0$, $D_s^+/D^+$ and $\Lambda_c^+/D_s^+$, to measure
various properties of low $p_T$ charm quark hadronization.
We argue that ratios $D_s^+/D^0$ and $2D_s^+/(D^0+D^+)$ as the function of transverse momentum can serve as good
probes for the dynamics of charm quark hadronization.
We further make predictions for the yields and $p_T$ spectra of identified charm hadrons
in central Pb + Pb collisions at $\sqrt{s_{NN}}= 2.76$ TeV.

\end{abstract}

\pacs{25.75.Ag, 25.75.Dw, 25.75.Gz, 25.75.-q}
\maketitle

\section{introduction}

Collisions of heavy nuclei with ultra-relativistic collision energies provide conditions for the creation of a new phase Quark Gluon Plasma (QGP) in the laboratory.
Hadrons containing charm quarks are powerful tools to study this high density and strong interacting QGP \cite{JPsisuprQGP,Cdiffusetherm,CCbarQGPtemp,DsprobeQGP}.
Measurements on charm hadrons have been executed at the BNL Relativistic Heavy Ion Collider (RHIC) \cite{D02005PRL,NcRHICphenix} and recently at the CERN Large
Hadron Collider (LHC) \cite{D0crosssec,DJHEP2012,Dv2,JPsiv2}.
With the increase of collision energy from RHIC to LHC, the production of charm quarks increases rapidly.
At LHC energies, charm quarks produced in QGP stage
become comparable to those produced in initial hard rescatterings \cite{csourcePLV1997,csourceJUp2010,csourceBWZ2008}.
In addition, re-interactions of charm quarks with partons in QGP also increase with collision energy due to the increasing temperature and energy density of
the medium produced in collisions \cite{csourcePLV1997,csourceJUp2010,csourceBWZ2008,charm-thermal}.
The hadronization of charm quarks is an important topic in charm field
since it plays a role of translating these initial or early effects into hadronic observables,
and has attracted lots of attentions \cite{DAAPLB2004,TYao2008PRC,RlamcD0com,Bccombi2013PRC,charmCombFrag,HJXu2013}.
Quark Combination Mechanism (QCM) is one of the effective phenomenological methods to describe charm quark
hadronization. It has shown its success early in studying flavor dependencies of open charm meson and baryon production in elementary hadronic
collisions \cite{Dpip1995,DLamcPLB2001,DpiN2002,DpiNpN2003,DLcpiNJPG2004} and has many applications in heavy ion collisions
recently \cite{DAAPLB2004,TYao2008PRC,RlamcD0com,Bccombi2013PRC}.

In this paper, we extend the quark combination model in Ref \cite{CEShao2009PRC,RQWang2012PRC} by including charm quarks to study the yield correlations
and $p_T$ dependence of charm hadrons in central Pb + Pb collisions at $\sqrt{s_{NN}}= 2.76$ TeV.
Hadron yield correlations, measured mainly by the ratios of yields of different hadrons, are one kind of effective probes for the mechanism 
of hadron production in high energy reactions \cite{RQWang2012PRC,thermal1995PLB,recom2004PRC,co2005PRC,Zimanyi2000PLB}. 
We study several kinds of two-particle yield ratios such as $D^{*+}/D^0$, $D_{s}^+/D^+$ 
and $\Lambda_c^+/D_s^+$ as well as four-particle yield correlations, by virtue of which we explore the properties of low  $p_T$ charm quark 
hadronization from different aspects.
Furthermore, we study $p_T$ spectra of various charm hadrons based on the explanation of the experimental data of various light and strange hadrons.
The $p_T$ dependence of charm meson ratios $D_s^+/D^0$ and $2D_s^+/(D^0+D^+)$ are especially selected to probe the dynamics of charm quark hadronization.

The rest of the paper is organized as follows. In Sec.~II, we systematically study the yield ratios and multiplicities of charm hadrons in the QCM.
In Sec.~III, we present the $p_T$ spectra of light, strange and charm hadrons as
well as the $p_T$ dependence of charm meson ratios $D_s^+/D^0$ and $2D_s^+/(D^0+D^+)$. Sec.~IV summaries our work.

\section{yield correlations and multiplicities of charm hadrons in the QCM}   \label{yield-GQC}

In this section, we extend the formulae of hadron yields in Ref \cite{RQWang2012PRC} to incorporating charm hadrons in the QCM.
We discuss several interesting yield ratios of charm hadrons which can reflect various properties in their production and
predict the midrapidity hadron yields in central Pb + Pb collisions at $\sqrt{s_{NN}}=2.76$ TeV.

\subsection{The formalism of hadron yield}

We start with a color-neutral quark-antiquark system with $N_{q_i}$ quarks of flavor $q_i$ and $N_{\bar q_i}$ antiquarks
of flavor $\bar q_i$.
These quarks and antiquarks hadronize via the quark combination.
The average numbers for the directly produced mesons $M_j$ and baryons $B_j$ are given by Ref \cite{RQWang2012PRC},
\begin{eqnarray}
 &&\overline{N}_{M_j}(N_{q_i},N_{\bar{q}_i})=\mathcal{P}_{M_j}\overline{N}_{M}(N_{q},N_{\bar{q}},\sqrt{s_{NN}})  = C_{M_j} \frac{N_{q_1 \bar{q}_2}}{N_{q \bar{q}}}
 \overline{N}_{M}(N_{q},N_{\bar{q}},\sqrt{s_{NN}}),     \label{eq:Mj-qi}      \\
%----------------------------------------------------------------------------------------------------
 &&\overline{N}_{B_j}(N_{q_i},N_{\bar{q}_i})= \mathcal{P}_{B_j} \overline{N}_{B}(N_{q},N_{\bar{q}},\sqrt{s_{NN}}) = N_{iter}C_{B_j}
 \frac{N_{q_1q_2q_3}}{N_{qqq}} \overline{N}_{B}(N_{q},N_{\bar{q}},\sqrt{s_{NN}}). ~~~~~~~  \label{eq:Bj-qi}
\end{eqnarray}
$\overline{N}_{M}$ and $\overline{N}_{B}$ are the average numbers of all mesons and all baryons produced in the system, and they are
functions of the total quark number $N_q=\sum_i N_{q_i}$ and antiquark number $N_{\bar q}=\sum_i N_{\bar q_i}$
due to the flavor-blind property of the strong interaction. Their properties are discussed in detail in Ref \cite{JS2013PRC}.
$\mathcal{P}_{M_j}$ is the production weight of a specific meson $M_j$ as a meson is formed.
It is determined by the probability of finding a specific $q_1\bar{q}_2$ pair in all $q\bar{q}$ pairs, ${N_{q_1 \bar{q}_2}}/{N_{q \bar{q}}}$,
and the flavor branch ratio of $M_j$ in the corresponding flavor-multiplets $C_{M_j}$,
where $N_{q_1 \bar q_2}= N_{q_1 }N_{\bar q _2}$ is the number of all the possible $q_1 \bar q_2$ pairs and $N_{q\bar q} =  N_{q }N_{\bar q}$ the number
of all the possible $q\bar{q}$ pairs.
Decomposition of $\mathcal{P}_{B_j}$ is similar to $\mathcal{P}_{M_j}$ except an extra $N_{iter}$ which stands for the number of possible
iterations of $q_1q_2q_3$. It is taken to be $1$, $3$, and $6$ for three identical flavor, two different flavor, and three different flavor cases, respectively.
$N_{q_1 q_2 q_3}$ is the number of all the possible $(q_1 q_2 q_3)$'s which satisfies
$N_{q_1 q_2 q_3}=N_{q_1}N_{q_2}N_{q_3}$ for $q_1\neq q_2 \neq q_3$, $ N_{q_1 q_2 q_3}=N_{q_1}(N_{q_1}-1)N_{q_3}$ for $q_1 = q_2 \neq q_3$
and $N_{q_1 q_2 q_3}=N_{q_1}(N_{q_1}-1)(N_{q_1}-2)$ for $q_1= q_2 = q_3$. $N_{qqq}$ is the number of all the possible $(qqq)$'s,
satisfying $N_{qqq}=N_{q}(N_{q}-1)(N_{q}-2)$. For flavor branch ratios $C_{M_j}$ and $C_{B_j}$ of charm hadrons, similar to light and strange hadrons, only
including $J^P=0^-$ and $1^-$ mesons and $J^P=\frac{1}{2}^+$ and $\frac{3}{2}^+$ baryons and using
the factors $R_{V/P}$ and/or $R_{O/D}$ to denote the relative production weight of hadrons with the
same flavor composition, we have
\begin{equation}
C_{M_j} =  \left\{
\begin{array}{ll}
{1}/{(1+R_{V/P})}~~~~~~~~   \textrm{for } J^P=0^-  \textrm{ charm mesons}  \\
{R_{V/P}}/{(1+R_{V/P})}~~~~         \textrm{for } J^P=1^-  \textrm{ charm mesons},
\end{array} \right.
\end{equation}
and
\begin{equation}
C_{B_j} =  \left\{
\begin{array}{ll}
{R_{O/D}}/{(1+R_{O/D})}~~~~   \textrm{for } J^P=({1}/{2})^+  \textrm{ charm baryons}  \\
{1}/{(1+R_{O/D})}~~~~~~~~         \textrm{for } J^P=({3}/{2})^+  \textrm{ charm baryons},
\end{array} \right.
\end{equation}
except that $C_{\Lambda_c^+}=C_{\Sigma_c^+}=C_{\Xi_c^0}=C_{\Xi_c^{'0}}=C_{\Xi_c^+}=C_{\Xi_c^{'+}}={R_{O/D}}/{(1+2R_{O/D})}$,
$C_{\Sigma_c^{*+}}=C_{\Xi_c^{*0}}=C_{\Xi_c^{*+}}={1}/{(1+2R_{O/D})}$,
and $C_{\Omega_{ccc}^{*++}}=1$.

For a reaction at a given energy, the average numbers of quarks of different flavors $\langle N_{q_i} \rangle$ and those of antiquarks of different
flavors $\langle N_{\bar{q}_i} \rangle$ are
fixed while $N_{q_i}$ and $N_{\bar{q}_i}$ follow a certain distribution.
In this work, we focus on the midrapidity region at so high LHC energy that the influence of net quarks can be ignored \cite{pikpPRL2012}.
We suppose a multinomial distribution
for both the numbers of $u$, $d$, $s$ and $c$ quarks at a given $N_q$ and the numbers of $\bar u$, $\bar d$, $\bar s$ and $\bar c$ at a given $N_{\bar{q}}$ with the prior
probabilities $p_u=p_d=p_{\bar u}=p_{\bar d}=1/(2+\lambda_s+\lambda_c)$, $p_s=p_{\bar s}=\lambda_s/(2+\lambda_s+\lambda_c)$,
$p_c=p_{\bar c}=\lambda_c/(2+\lambda_s+\lambda_c)$. Here,
we introduce two factors $\lambda_s$ and $\lambda_c$ to denote the production suppression of strange quarks and charm quarks, respectively.
Averaging over this distribution, Eqs.~(\ref{eq:Mj-qi}-\ref{eq:Bj-qi}) become
\begin{eqnarray}
 \overline{N}_{M_j}(N_{q},N_{\bar q},\sqrt{s_{NN}})
 &=&C_{M_j}\ p_{q_{1}}p_{\bar{q}_{2}} \overline{N}_{M}(N_{q},N_{\bar q},\sqrt{s_{NN}}),        \label{eq:Mj-q} \\
%----------------------------------------------------------------------------------------------------
 \overline{N}_{B_j}(N_{q},N_{\bar q},\sqrt{s_{NN}})
 &=&N_{iter}C_{B_j}\ p_{q_{1}}p_{q_{2}}p_{q_{3}} \overline{N}_{B}(N_{q},N_{\bar q},\sqrt{s_{NN}}). ~~~~~~  \label{eq:Bj-q}
\end{eqnarray}

Since yields measured by experiments usually cover only a limited kinematic region, we should also consider the fluctuations of $N_{q}$ and $N_{\bar q}$
in this kinematic region. By averaging over this fluctuation distribution of $N_{q}$ and $N_{\bar q}$ with the
fixed $\langle N_{q} \rangle$ and $\langle N_{\bar q} \rangle$, we have
\begin{eqnarray}
 \langle N_{M_j}\rangle(\langle N_{q}\rangle,\langle N_{\bar q}\rangle,\sqrt{s_{NN}})
  &=&C_{M_j}\ p_{q_{1}}p_{\bar{q}_{2}}\langle N_{M}\rangle(\langle N_{q}\rangle,\langle N_{\bar q}\rangle,\sqrt{s_{NN}}),
 \label{eq:NMj_aver}     \\
%----------------------------------------------------------------------------------------------------
 \langle N_{B_j}\rangle(\langle N_{q}\rangle,\langle N_{\bar q}\rangle,\sqrt{s_{NN}})
 &=&N_{iter}C_{B_j}\ p_{q_{1}}p_{q_{2}}p_{q_{3}}\langle N_{B}\rangle(\langle N_{q}\rangle,\langle N_{\bar q}\rangle,\sqrt{s_{NN}}), ~~~~~~
 \label{eq:NBj_aver}
\end{eqnarray}
where $\langle N_{M}\rangle$ and $\langle N_{B}\rangle$ stand for the average total number of the mesons and that of the baryons
produced in the combination process.

Finally we incorporate the decay contribution of short lifetime hadrons to obtain
\begin{equation}
\langle N_{h_j}^f\rangle = \langle N_{h_j}\rangle+ \sum_{i\not=j} Br(h_i\to h_j) \langle N_{h_i}\rangle,
\end{equation}
where the superscript $f$ denotes final hadrons. Decay branch ratio $Br(h_i\to h_j)$ is given by the Particle Data Group \cite{PDG2012}.

\subsection{Yield correlations of charm hadrons}

From the above discussion, we get that the yields of charm hadrons depend on several physical parameters, e.g., $R_{V/P}$, $\lambda_s$ and $\lambda_c$, etc.,
which can be seen in Table \ref{tab_yields} for
a clear presentation.
These parameters, however, can be measured experimentally by the yield ratios between different hadrons.
We mainly propose the following four types of particle ratios.

\begin{table}[htbp]
\renewcommand{\arraystretch}{1.5}
 \centering
 \caption{Yields of the directly produced charm hadrons and those including the Strong and ElectroMagnetic (S\&EM) decay contributions.}
  \begin{tabular}{p{40pt}p{80pt}p{80pt}}
    \hline\hline
    hadrons      & directly produced  & with S\&EM decays   \\
    \hline
    $D_s^{+}$      &$\frac{1}{1+R_{V/P}} p_cp_{\bar s} \langle N_M\rangle$
                &$p_cp_{\bar s} \langle N_M\rangle$  \\

  $\Lambda_c^+$    &$\frac{6R_{O/D}}{1+2R_{O/D}} p_{u}^2p_c \langle N_B\rangle$
                &$12p_{u}^2p_c \langle N_B\rangle$  \\

   $D^{0}$        &$\frac{1}{1+R_{V/P}} p_cp_{\bar u} \langle N_M\rangle$
                &$\frac{1+1.677R_{V/P}}{1+R_{V/P}} p_cp_{\bar u} \langle N_M\rangle$ \\

  $D^{+}$        &$\frac{1}{1+R_{V/P}} p_cp_{\bar u} \langle N_M\rangle$
                &$\frac{1+0.323R_{V/P}}{1+R_{V/P}} p_cp_{\bar u} \langle N_M\rangle$ \\

  $D^{*+}$       &$\frac{R_{V/P}}{1+R_{V/P}} p_cp_{\bar u} \langle N_M\rangle$
                &$\frac{R_{V/P}}{1+R_{V/P}} p_cp_{\bar u} \langle N_M\rangle$ \\

  $D^{*0}$       &$\frac{R_{V/P}}{1+R_{V/P}} p_cp_{\bar u} \langle N_M\rangle$
                &$\frac{R_{V/P}}{1+R_{V/P}} p_cp_{\bar u} \langle N_M\rangle$  \\

  $J/\Psi$       &$\frac{R_{V/P}}{1+R_{V/P}} p_cp_{\bar c} \langle N_M\rangle$
                &$\frac{R_{V/P}}{1+R_{V/P}} p_cp_{\bar c} \langle N_M\rangle$  \\
    \hline \hline
 \end{tabular}     \label{tab_yields}
\end{table}

First, ratios of $D^*$ to $D$ mesons that can reflect $R_{V/P}$, such as
\begin{eqnarray}
   \frac{2\langle N_{D^{*+}}^{f}\rangle}{\langle N_{D^0}^{f}\rangle + \langle N_{D^+}^{f}\rangle}=\frac{R_{V/P}}{1+R_{V/P}},  ~~~~~~~~
   \frac{\langle N_{D^{*+}}^{f}\rangle}{\langle N_{D^0}^{f}\rangle}=\frac{R_{V/P}}{1+1.677R_{V/P}},  ~~~~~~~~
   \frac{\langle N_{D^{*+}}^{f}\rangle}{\langle N_{D^+}^{f}\rangle}=\frac{R_{V/P}}{1+0.323R_{V/P}}.    \label{eq:Dratio}
\end{eqnarray}
By the measurement of these ratios we can quantify $R_{V/P}$, which is helpful for the understanding of the effects of spin interactions
during hadronization.
Besides, we can also gain the information of $R_{V/P}$ by a yield hierarchy among $D$ mesons. From the results in Table \ref{tab_yields}, we have
\begin{eqnarray}
  \langle N_{D^{0}}^{f}\rangle \colon \langle N_{D^{*+}}^{f}\rangle \colon \langle N_{D^{+}}^{f}\rangle
  =\frac{1+1.677R_{V/P}}{1+R_{V/P}}:\frac{R_{V/P}}{1+R_{V/P}}:\frac{1+0.323R_{V/P}}{1+R_{V/P}},   \label{eq:Dhierarchy}
\end{eqnarray}
and then we get $\langle N_{D^{0}}^{f}\rangle > \langle N_{D^{*+}}^{f}\rangle > \langle N_{D^{+}}^{f}\rangle$ when $R_{V/P}>1.477$
while $\langle N_{D^{0}}^{f}\rangle > \langle N_{D^{+}}^{f}\rangle > \langle N_{D^{*+}}^{f}\rangle$ when $R_{V/P}<1.477$.
Since the mass discrepancy between $D^{*}$ and $D$ is much smaller than that in light and strange hadrons, one could expect $D^{*}$ is not much suppressed
relative to $D$ and $R_{V/P}$ might approach 3 by counting the spin degree of freedom. 
We note that $R_{V/P}$ has been observed to approximately equal to 3 in charm sector in $pp$ collisions at LHC \cite{gammasPP7}.
In this case  Eqs. (\ref{eq:Dratio}-\ref{eq:Dhierarchy}) are
easily tested by the experiments.

Second, ratios between $D_s$ and $D$ mesons that can reflect the strangeness suppression factor $\lambda_s$, e.g.,
\begin{eqnarray}
  \frac{2\langle N_{D_s^{+}}^{f}\rangle}{\langle N_{D^0}^{f}\rangle+\langle N_{D^+}^{f}\rangle}=\lambda_s, ~~~~~~~~
  \frac{\langle N_{D_s^{+}}^{f}\rangle}{\langle N_{D^0}^{f}\rangle}=\frac{1+R_{V/P}}{1+1.677R_{V/P}} \lambda_s, ~~~~~~~~
  \frac{\langle N_{D_s^{+}}^{f}\rangle}{\langle N_{D^+}^{f}\rangle}=\frac{1+R_{V/P}}{1+0.323R_{V/P}} \lambda_s.  \label{eq:RDSD}
\end{eqnarray}
In the QCM, a charm quark captures a light or a strange antiquark to form a $D$ meson or a $D_s$ meson, so ratios of these $D_s$ to $D$ mesons
carry the strangeness information of the production zone of these open charm mesons.
By virtue of the strangeness, these ratios can effectively probe the hadronization environment of charm quarks,
i.e., inside or outside the QGP. It is known that in
relativistic heavy ion collisions the produced bulk quark matter has the saturated strangeness which is about 0.40-0.45, while for the small partonic system such as that created in $pp$ collisions the strangeness is only about 0.3 \cite{gammasPP7}. Therefore, if these ratios measured by experiments tend to the
saturated value, this means
charm quarks mainly hadronize in the QGP, otherwise most of them pass through the QGP and hadronize outside.

Third, ratios that reflect the suppression of charm quarks $\lambda_c$, e.g.,
\begin{eqnarray}
  \frac{\langle N_{J/\psi}^{f}\rangle}{\langle N_{D^{*+}}^{f}\rangle}
  =\frac{\langle N_{J/\psi}^{f}\rangle}{\langle N_{D^{*0}}^{f}\rangle}  =\lambda_c,~~~~~~
  \frac{\langle N_{\Lambda_c^+}^{f}\rangle}{\langle N_p^{f}\rangle}
  =\frac{\langle N_{\Lambda_c^+}^{f}\rangle}{\langle N_n^{f}\rangle} =3\lambda_c,
\end{eqnarray}
where we use $\langle N_p^{f}\rangle=\langle N_n^{f}\rangle=4p_u^3\langle N_B\rangle$ obtained in our
previous work \cite{RQWang2012PRC}.
These ratios provide the direct measurement of the production suppression of charm quarks relative to light quarks.

Fourth, ratios that reflect the baryon-meson competition in the charm sector, e.g.,
\begin{eqnarray}
  \frac{\langle N_{\Lambda_c^+}^{f}\rangle}{\langle N_{D^0}^{f}\rangle+\langle N_{D^+}^{f}\rangle}
  =\frac{6}{2+\lambda_s+\lambda_c}\frac{\langle N_B\rangle}{\langle N_M\rangle}, ~~~~~~~~~~~~~~~~~~
  \frac{\langle N_{\Lambda_c^+}^{f}\rangle}{\langle N_{D_s^+}^{f}\rangle}
  =\frac{12}{\lambda_s(2+\lambda_s+\lambda_c)}\frac{\langle N_B\rangle}{\langle N_M\rangle}.
\end{eqnarray}
Because of small decay contaminations, these ratios provide a clean measurement of baryon and meson production competition.

In addition, we can build more sophisticated combinations of the average yields of different hadrons such as
\begin{equation}
\frac{\langle N_{D_s^+}^{f}\rangle \langle N_{\Xi^-}^{f}\rangle}{(\langle N_{D^0}^{f}\rangle+\langle N_{D^+}^{f}\rangle) \langle N_{\Omega^-}^{f}\rangle}=\frac{3}{2}, ~~~~
\frac{\langle N_{\Lambda_c^+}^{f}\rangle \langle N_{D^{*+}}^{f}\rangle}{\langle N_{p}^{f}\rangle \langle N_{J/\psi}^{f}\rangle}=3,
\end{equation}
where we use $\langle N_{\Xi^-}^{f}\rangle=3p_up_s^2\langle N_B\rangle$ and $\langle N_{\Omega^-}^{f}\rangle=p_s^3\langle N_B\rangle$ \cite{RQWang2012PRC}.
These two relations are independent of the above parameters.  
Therefore, they are characteristics of the QCM and can be used for the first test of the QCM.

\subsection{Predictions for charm hadron yields}

Now we turn to the predictions of the midrapidity yields $dN/dy$ of various hadrons in central Pb + Pb collisions at $\sqrt{s_{NN}}= 2.76$ TeV.
We first fix the necessary inputs.
The rapidity density of all quarks $dN_{q}/dy$ is fixed to be 1731 by fitting
the measured pseudorapidity density of charged particles \cite{Nch1601,taoyaothesis}.
The strangeness suppression factor $\lambda_s$ is taken to be the saturation value $0.41$ and then we get the rapidity density
of strange quarks $\frac{dN_{s}}{dy}=\frac{\lambda_s}{2+\lambda_s+\lambda_c}\frac{dN_{q}}{dy} \approx \frac{\lambda_s}{2+\lambda_s}\frac{dN_{q}}{dy}$.
Charm quark number which is related to the suppression factor $\lambda_c$ by $\frac{dN_{c}}{dy}=\frac{\lambda_c}{2+\lambda_s+\lambda_c}\frac{dN_{q}}{dy}$ is
the key input for predictions of charm hadron yields.

The production of charm quarks at LHC mainly comes from two stages of the collision: initial hard-parton scatterings and thermal partonic interactions in the QGP.
Here we first give an estimation of charm quark number $dN_c/dy$ from initial hard scatterings by extrapolating $pp$ reaction data at LHC,
 \begin{equation}
  \frac{dN^{PbPb}_{c}}{dy}=\frac{1}{R}\frac{dN^{PbPb}_{D^0}}{dy}=\frac{1}{R}<T_{AA}>\frac{d\sigma^{pp}_{D^0}}{dy}=21\pm6.
 \end{equation}
Here $R=0.54\pm0.05$ is the branch ratio of charm quarks into final $D^0$ mesons measured in $e^+e^-$ reactions \cite{D02005PRL}.
$<T_{AA}>=26.4\pm0.5~\mathrm{mb}^{-1}$ is the average nuclear overlap function calculated with the Glauber model \cite{HpmPLB2011}.
The cross section of $D^0$ is $\frac{d\sigma_{D^0}^{pp}}{dy}=0.428\pm0.115$ mb in $pp$ reactions at $\sqrt{s}= 2.76$ TeV \cite{D0crosssec}.

Subsequent QGP evolution stage can increase the charm quark number but the enhancement is sensitive to the evolution details
such as initial temperature, charm quark mass, evolution time, etc,
which
lead to that the present predictions of charm quark number in literatures have a large uncertainties \cite{csourcePLV1997,csourceBWZ2008,csourceJUp2010}.
Here, we consider the largest uncertainty predicted by Ref \cite{csourcePLV1997} which is up to $100\%$, i.e., $\frac{dN_c}{dy}=21^{+27}_{-6}$,
to predict charm hadron yields and their uncertainties.
The predicted yields of various charm hadrons as well as light and strange hadrons are shown in Table \ref{tab:Nhadron},
and the data are from Refs \cite{pikpPRL2012,Ks0lam2013,xiome2013}.
Here, the remaining parameter $R_{V/P}$ is taken to be 3 by the spin counting and $R_{O/D}$ is taken to be 2, the same as in the light and strange sectors.
The results for light and strange hadrons are consistent with the available data within the experimental uncertainties.

Note that the current prediction of $\phi$ mesons might be higher than the future datum.
This is because that in this paper we only consider the production of SU(4) hadrons within the quark model.
In fact, various multi-quark states and/or exotic states beyond the quark model are also produced in high energy reactions.
There are two exotic hadrons, i.e. $f_0(980)$ and $a_0(980)$, that are mostly relevant to the $\phi$ production abundance.
The masses of these two hadrons are close to $\phi$ and they also probably have the large $s\bar{s}$ content according to the newest PDG \cite{PDG2012}.
Including these two hadrons in our model will consume $s\bar{s}$ pairs in the system and introduce the competition against the $\phi$ formation. 
According to the experimental data in $e^+e^-$ annihilations \cite{PDG2012} and our previous study at top SPS \cite{CEShao2009PRC}, 
we estimate the practical $\phi$ yield in central Pb + Pb collisions at $\sqrt{s_{NN}}=2.76$ TeV might be about 14, i.e. 
only a half of current prediction.

\begin{table*}[htbp]
\renewcommand{\arraystretch}{1.5}
\caption{The rapidity densities $dN/dy$ of identified hadrons at $dN_c/dy=21^{+27}_{-6}$. The data are from Refs \cite{pikpPRL2012,Ks0lam2013,xiome2013}.}
\begin{tabular}{p{40pt}p{80pt}p{60pt}p{40pt}p{80pt}p{40pt}}
\toprule
Hadron   & Data      & Results       & Hadron  & Data      & Results  \\
\colrule
$\pi^+$  &$733\pm54$ &$741^{+5}_{-24}$  &$K^+$    &$109\pm9$  &$110^{+1}_{-4}$   \\
$K_S^{0}$ &$110\pm10$  &$105^{+1}_{-3}$   &$\phi$   &---        &$29.3^{+0.2}_{-0.9}$   \\
$p$      &$34\pm3$   &$32^{+1}_{-2}$     &$\Lambda$ &$26\pm3$  &$25\pm1$   \\
$\Xi^-$  &$3.34\pm0.06\pm0.24$ &$4.01^{+0.04}_{-0.19}$ &$\Omega^-$ &$0.58\pm0.04\pm0.09$ &$0.55^{+0.01}_{-0.03}$  \\
\colrule
$D^+$    &---        &$3.39^{+4.24}_{-0.96}$     &$D^0$    &---       &$10.4^{+13.0}_{-3.0}$     \\
$D^{*+}$ &---        &$5.16^{+6.46}_{-1.46}$     &$D_s^+$  &---       &$2.82^{+3.53}_{-0.80}$     \\
$J/\Psi$ &---        &$0.15^{+0.65}_{-0.07}$     &$\Lambda_c^+$  &---   &$2.82^{+3.43}_{-0.79}$    \\
\botrule
\end{tabular} \label{tab:Nhadron}
\end{table*}

\section{$p_T$ dependence of light, strange and charm hadrons}

$p_T$ spectra of identified hadrons, involving
the overall multiplicity as well as the $p_T$ distribution for each of particle species,
can provide more explicit insights into quark hadronization mechanisms and detailed information on the hot and dense bulk matter created
in relativistic heavy ion collisions. In the QCM, $p_T$ spectra of various light and strange hadrons can be
systematically calculated with the given $p_T$ distributions of light and strange quarks. 
This is one of the main features and advantages of the QCM and has tested against the experimental data at
RHIC \cite{recom2004PRC,recom2007PRC,co2005PRC,co2006PRC,YFWang2008CPC,SDQCM2012PRC}. Here, we test the QCM
in explaining $p_T$ spectra of identified hadrons at LHC by using a specific Quark Combination Model developed by Shandong Group \cite{QBXie1988PRC,FLShao2005PRC,CEShao2009PRC}.
For the input distributions of
light and strange quarks, we use a two-component parameterized pattern:
$d^2N_{q_i}/(p_Tdp_Tdy) \propto \exp(-\sqrt{p_T^2+m_{q_i}^2}/T_{q_i})+R_{q_i}(1+\frac{p_T}{5GeV})^{-S_{q_i}}$,
the exponential item for thermal quarks at low $p_T$ and
power-law item for shower quarks with high $p_T$. $m_{q_i}$ is the constituent quark mass, and $m_u=m_d=0.34$ GeV and $m_s=0.5$ GeV.
Parameters ($T_{q_i}, S_{q_i},R_{q_i}$ ) for light and strange quarks are extracted by fitting the data of $p_T$ spectra of $\pi^++\pi^-$
and $\Lambda$, respectively. Their values are (0.31 GeV, 8.0, 0.02) for light quarks and (0.36 GeV, 8.6, 0.04) for strange quarks.

\begin{figure}[htbp]
\centering
 \includegraphics[width=0.95\linewidth]{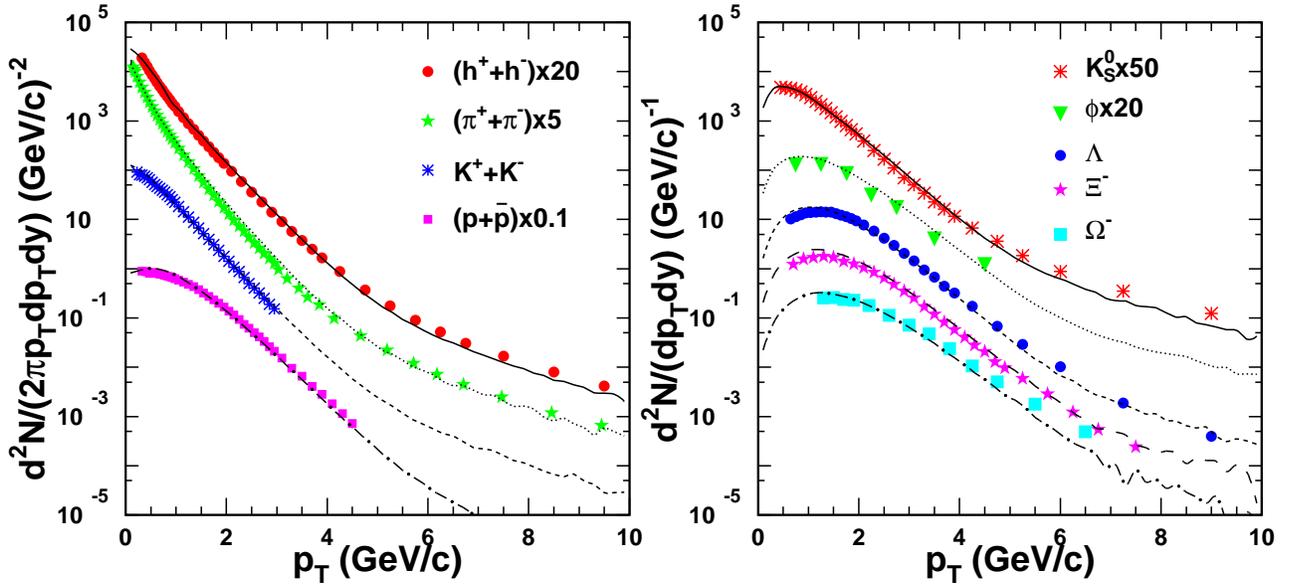}\\
 \caption{(Color online) $p_T$ distributions of light and strange hadrons in central
Pb + Pb collisions at $\sqrt{s_{NN}}=2.76$ TeV.
 The symbols are the experimental data from
 Refs \cite{Ks0lam2013,xiome2013,HpmPLB2011,pikpPRL2012,inverseRAA,phipt}. Charged particles $h^{\pm}$ are measured in midpsudorapidity,
and identified hadrons are in midrapidity. Note that the data of $\pi^++\pi^-$ and $\Lambda$ are used
to extract the $p_T$ spectra of light and strange quarks, respectively.}   \label{centralHPT}
\end{figure}

Fig.~\ref{centralHPT} shows the $p_T$ spectra of $h^{\pm}$, $\pi^\pm$, $K^\pm$,
$K_S^0$, $\phi$, $p$, $\bar p$, $\Lambda$, $\Xi^-$ and $\Omega^-$ in central Pb + Pb collisions at $\sqrt{s_{NN}}=2.76$ TeV and the calculated results are
consistent with the experimental data \cite{Ks0lam2013,xiome2013,HpmPLB2011,pikpPRL2012,inverseRAA,phipt}.
This exhibits the validity of the QCM in light and strange sectors at LHC.
We see that the exponential domain of meson spectra expands to about 4 GeV/c and that of baryons expands to about 6 GeV/c.
Compared with the data at RHIC \cite{Hpt2006PRL}, the exponential domain extends over about 2 GeV/c.

\begin{figure}[htbp]
\centering
 \includegraphics[width=0.9\linewidth]{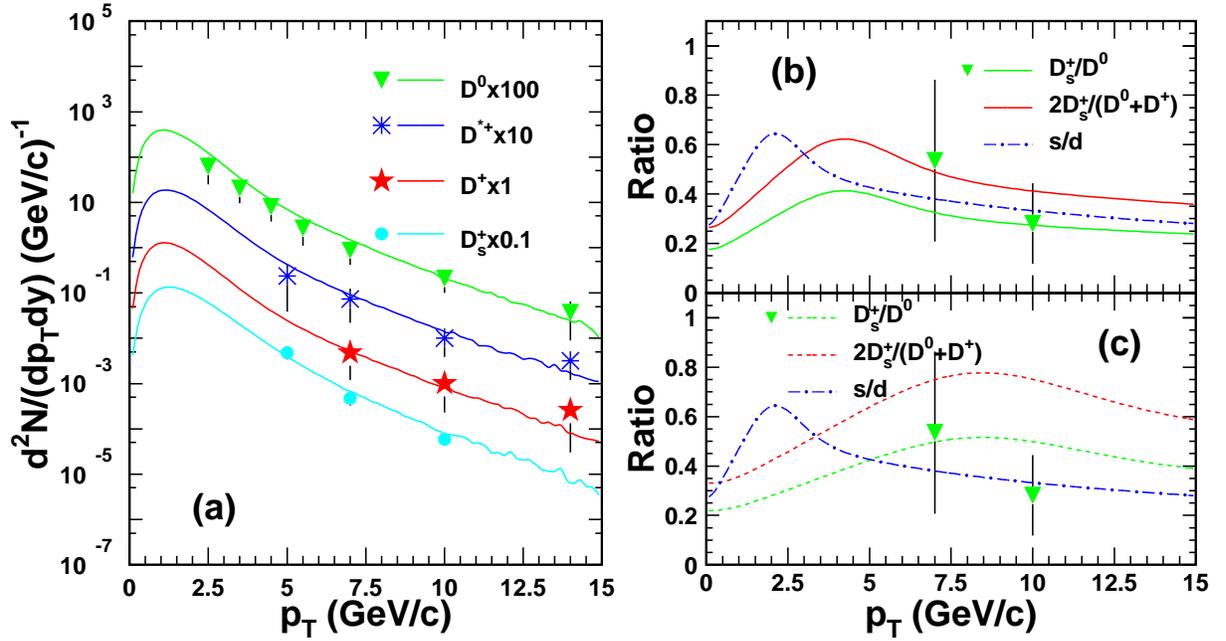}\\
 \caption{(Color online) (a) $p_T$ distributions of charm mesons and (b) charm meson ratios in equal $p_T$ combination scenario
 and (c) charm meson ratios in equal velocity combination scenario at midrapidity in central
Pb + Pb collisions at $\sqrt{s_{NN}}=2.76$ TeV.
 The filled symbols are the experimental data from Refs \cite{DJHEP2012,Dsp}. Note that the data of $D^0$ are used
to extract the $p_T$ spectra of charm quarks.}
 \label{centralDPT}
\end{figure}

Based on the performance of the QCM in light and strange sectors, we turn to the charm sector.
The $p_T$ distribution of charm quarks is the relevant input for the predictions of various charm hadrons. We extract it from the data of $D^0$ as
$d^2N_{c}/(p_Tdp_Tdy) \propto (p_T+0.4GeV)^{2}(1+\frac{p_T}{2.2GeV})^{-8}$. This input is within the theoretical prediction range of charm quark spectrum
in Ref \cite{charm-pt}.
Fig.~\ref{centralDPT} (a) shows the calculated $p_T$ spectra of $D$ mesons at midrapidity in central Pb + Pb collisions at $\sqrt{s_{NN}}=2.76$ TeV,
from which one can see that within the error uncertainties the results agree with the data from Refs \cite{DJHEP2012,Dsp}.

We argue that the $p_T$ dependence of charm meson ratios $D_s^+/D^0$ and $2D_s^+/(D^0+D^+)$ can help explore the dynamical part of charm quark hadronization.
These two ratios, as
are shown in Eq.~(\ref{eq:RDSD}), are proportional to the strangeness of the production regions of charm mesons.
The dot-dashed lines in Fig.~\ref{centralDPT} (b) and (c) show the $p_T$ dependence of the strangeness, i.e., the ratio of
the previously extracted strange quark $p_T$ spectrum to down quark $p_T$ spectrum.
It rises rapidly at low $p_T$ and reaches the peak at $p_T\approx 2$ GeV and then it decreases to a stable small value.
$D_s^+/D^0$ and $2D_s^+/(D^0+D^+)$ as the function of $p_T$ should follow similar sharps by stretching out the $p_T$ axis,
but the peak position of the ratios will reflect the detailed combination dynamics of charm hadron formation.
The solid lines and dashed lines in panel (b) and (c), respectively, show our predictions of these two ratios in two different combination
dynamics, i.e. equal $p_T$ or equal velocity combination, which can not be effectively discriminated in the experimental data of single particle spectra discussed above.
Panel (b) shows the results of a charm quark capturing an antiquark with an almost equal $p_T$ to form a meson. The resulting $D_s^+/D^0$
and $2D_s^+/(D^0+D^+)$ reach the peak at $p_T \approx 4$ GeV, and decrease to the low values at high $p_T$.
Panel (c) represents the results in case of a charm quark capturing an antiquark with an almost equal velocity to form a meson.
In this case, the $p_T$ of charm quark is about triple of that of the antiquark in forming a $D$ meson due to their triple difference in mass.
The resulting $D_s^+/D^0$ and $2D_s^+/(D^0+D^+)$ will arrive at the peak at $p_T \approx 8$ GeV, which are quite different from those in panel (b).
The future experimental data at LHC can check these two different combination scenarios and provide deep insights into the hadronization
dynamics of charm quarks in ultra-relativistic heavy ion collisions.

\section{summary}

We have studied in the QCM the yield correlations and $p_T$ spectra of charm hadrons in central Pb + Pb collisions
at $\sqrt{s_{NN}}= 2.76$ TeV.
Yields of various charm hadrons are found to have a series of interesting correlations.
Several types of yield ratios were proposed to quantify these correlations and to
measure the properties of
charm quark hadronization from different aspects.
In addition, we used the $p_T$ spectra of light and
strange quarks extracted from the data of $\pi^++\pi^-$ and $\Lambda$, only two inputs, to systematically explain the midrapidity data of $p_T$ spectra for
$h^{\pm}$, $K^{\pm}$, $K^{0}_{S}$, $\phi$, $p$, $\bar p$, $\Xi^-$ and $\Omega^-$.
We further calculated the $p_T$ spectra of open charm mesons and found that the results agree with the available experimental data.
Ratios $D_s^+/D_{0}$ and $2D_s^+/(D^{0}+D^+)$ as the function of $p_T$ are identified as good probes for the hadronization dynamics of charm quarks, and
we made predictions in two different combination scenarios for the comparison with the future experimental data.

\section*{Acknowledgements}

The authors thank Z. T. Liang, Q. B. Xie, Q. Wang, H. J. Xu, W. Wang and the members of the particle theory group of Shandong University for helpful discussions.
This work is supported in part by the National Natural Science Foundation of China under grant 11175104, 11305076, 11247202, and
by the Natural Science Foundation of Shandong Province, China under grant ZR2011AM006, ZR2012AM001.

\end{document}